\documentclass[aps,pra,reprint,showpacs,floatfix,longbibliography]{revtex4-1}
\usepackage{graphicx, amsmath, amssymb}
\usepackage{hyperref}
\usepackage[caption=false]{subfig}
\allowdisplaybreaks[1] 

\begin{document}


\newcommand{\fref}[1]{Fig.~\ref{#1}}
\newcommand{\Tr}{\text{Tr}}
\newcommand{\zyz}{$ZY\!Z$ }


\title{Implementing Quantum Gates Using Length-3 Dynamic Quantum Walks}

\author{Ibukunoluwa A.~Adisa}
	\email{ibukunoluwaadisa@creighton.edu}
	\affiliation{Department of Physics, Creighton University, 2500 California Plaza, Omaha, NE 68178}

\author{Thomas G.~Wong}
	\email{thomaswong@creighton.edu}
	\affiliation{Department of Physics, Creighton University, 2500 California Plaza, Omaha, NE 68178}

\begin{abstract}
	It is well-known that any quantum gate can be decomposed into the universal gate set $\{ T, H, \text{CNOT} \}$, and recent results have shown that each of these gates can be implemented using a dynamic quantum walk, which is a continuous-time quantum walk on a sequence of graphs. This procedure for converting a quantum gate into a dynamic quantum walk, however, can result in long sequences of graphs. To alleviate this, in this paper, we develop a length-3 dynamic quantum walk that implements any single-qubit gate. Furthermore, we extend this result to give length-3 dynamic quantum walks that implement any single-qubit gate controlled by any number of qubits. Using these, we implement Draper's quantum addition circuit, which is based on the quantum Fourier transform, using a dynamic quantum walk.
\end{abstract}

\maketitle


\section{Introduction}

Despite the ubiquity of quantum circuits \cite{nielsen2002quantum}, other models of quantum computing exist. Quantum walks are an example. Also known as quantum random walks, they are the quantum versions of classical random walks, where a quantum particle evolves as a superposition over the vertices of a graph by moving along its edges. For a history of quantum walks, see Kempe's review \cite{Kempe2003}. In this paper, we focus on continuous-time quantum walks, which were first proposed by Farhi and Gutmann as a means of exploring decision trees \cite{Farhi1998}. In a continuous-time quantum walk, the evolution of the walker is governed by Schr\"odinger's equation,
\begin{equation}
	\label{schrodinger}
	i \frac{d}{dt}\left|\psi\left(t\right)\right\rangle = \hat{H} \left|\psi\left(t\right)\right\rangle,
\end{equation}
where we have set $\hbar = 1$, and $\hat{H}$ is an appropriate Hamiltonian that respects the graph on which the walk occurs. For the rest of the paper, we will refer to these simply as ``quantum walks,'' dropping the ``continuous-time'' adjective.

Many quantum algorithms are naturally framed as quantum walks. Examples include algorithms that traverse glued binary trees exponentially faster than classical computers \cite{childs2003exponential}, evaluate boolean expressions \cite{farhi2007quantum}, and search spatial regions \cite{childs2004spatial}. Childs \cite{childs2009universal} showed that quantum walks are universal for quantum computing, so any quantum circuit can be converted into a quantum walk. A ``rail'' or path of vertices was used for each computational basis state, with the quantum walk moving along the rails. By adding branches off these rails, adding connections between the rails, and swapping rails, Childs was able to implement the universal set of quantum gates $\{ T, H, \text{CNOT} \}$, where $T$ is the fourth root of the $Z$ gate, $H$ is the Hadamard gate, and CNOT is the controlled-NOT gate. Then, any quantum gate can be implemented by a quantum walk by decomposing it in terms of $T$, $H$, and CNOT gates, and then implementing each gate using Childs' gadgets. If the quantum gate acts on $n$ qubits, which has $N = 2^n$ computational basis states, then $N$ rails are needed for the quantum walk. Since each rail can contain many vertices, the total number of vertices for the quantum walk is larger than $N$.

Underwood and Feder \cite{Underwood2010} also explored the connection between quantum circuits and quantum walks. They again used the rail encoding, where each computational basis state corresponds to one rail, but now they permitted the edges to be weighted and to also change at discrete times. Thus, their approach can be thought of as a sequence of weighted graphs. By doing this appropriately, they were able to implement the $\sqrt{Z}$ gate, the single-qubit gate that rotates about the $x$-axis of the Bloch sphere by $2\sqrt{3}\pi$, and the controlled-$Z$ gate, and altogether, these three states form a universal set. Again, since the rail encoding was used, more than $N$ vertices are needed to implement a quantum circuit with $N$ computational basis states. 

Recently, Herrman and Humble \cite{herrman2019continuous} proposed a third method for implementing quantum circuits using quantum walks. Their approach abandoned the rail encoding, using instead a single vertex to encode each computational basis state. The edges of the graph were unweighted, and the edges were allowed to change at discrete times. Thus, the quantum walk occurred on a sequence of graphs, which they called a dynamic graph. They showed how the universal set of quantum gates $\{ T, H, \text{CNOT} \}$ could be implemented using quantum walks on dynamic graphs. In their construction, isolated vertices always had self-loops, and as a result, some of their implementations used ancillary vertices, meaning there could be more vertices than the number of computational basis states. Soon after, however, \cite{wong2019isolated} removed the need for ancillary vertices by permitting isolated vertices to be looped or loopless. With this simplification, a quantum circuit on $N$ computational basis states only takes $N$ vertices. In all of these results showing that quantum walks are universal for quantum computing, the Hamiltonian equalled the adjacency matrix of the graph, i.e.,
\[ H = A, \]
where $A_{ij} = 1$ if vertices $i$ and $j$ are adjacent (or in the case of \cite{Underwood2010}, $A_{ij}$ equaled the weight of the edge), and $A_{ij} = 0$ otherwise. We use this same Hamiltonian in this paper.

\begin{figure} 
\begin{center}
	\includegraphics{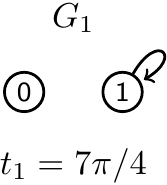} \quad
	\includegraphics{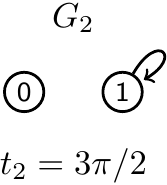} \quad
	\includegraphics{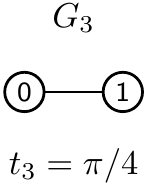} \quad
	\includegraphics{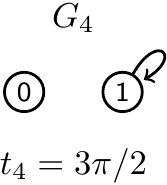}
	\vspace{0.2in}
    
	\includegraphics{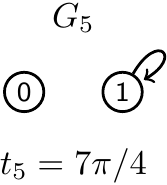} \quad
	\includegraphics{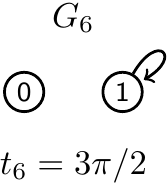} \quad
	\includegraphics{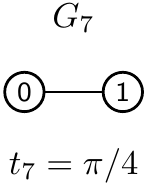} \quad
	\includegraphics{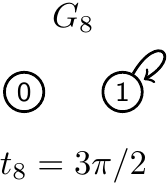}
	\vspace{0.2in}

	\includegraphics{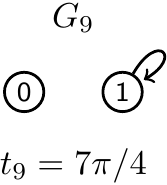} \quad
	\includegraphics{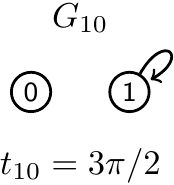} \quad
	\includegraphics{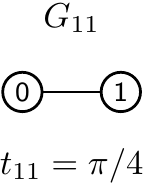} \quad
	\includegraphics{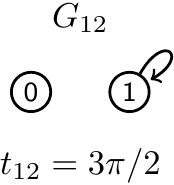}
	\vspace{0.2in}
    
	\includegraphics{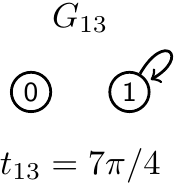} \quad
	\includegraphics{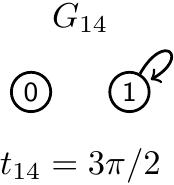} \quad
	\includegraphics{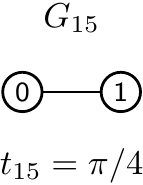} \quad
	\includegraphics{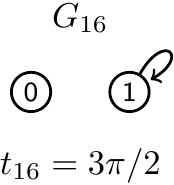}
	\vspace{0.2in}
    
	\includegraphics{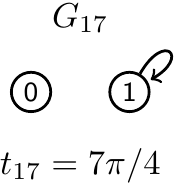} \quad
	\includegraphics{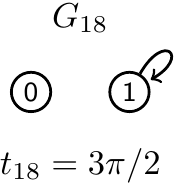} \quad
	\includegraphics{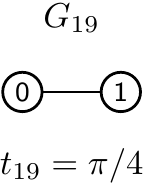} \quad
	\includegraphics{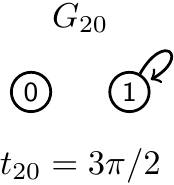}
	\vspace{0.2in}
    
	\includegraphics{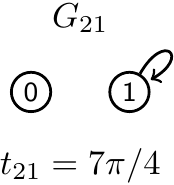} \quad
	\includegraphics{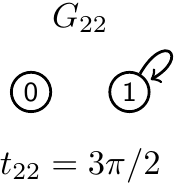} \quad
	\includegraphics{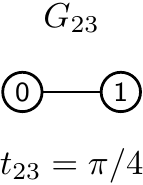} \quad
	\includegraphics{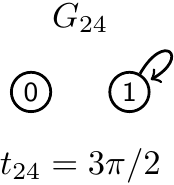}
    
	\caption{\label{HT6} A dynamic graph that implements $(HT)^{6}$.}
\end{center}
\end{figure}

For example, consider the following single-qubit gate:
\[ U = \frac{1}{4\sqrt{2}} \begin{pmatrix}
	3 + (1+\sqrt{2})i & 3-\sqrt{2} + (1+2\sqrt{2})i \\
	1+2\sqrt{2} + (3-\sqrt{2})i & -1-\sqrt{2} - 3i \\
\end{pmatrix}. \]
Following \cite{wong2019isolated}, to implement this quantum gate as a quantum walk on a dynamic graph, we first decompose the gate in terms of $T$, $H$, and CNOT gates, e.g., by using the Solovay-Kitaev theorem \cite{dawson2005solovay}. We find that
\[ U = HTHTHTHTHTHT = (HT)^{6}. \]
Then, we implement each of the $T$ and $H$ gates using results from \cite{wong2019isolated}. Each $T$ gate takes one graph and each Hadamard gate takes three graphs, for a total of twenty-four graphs. This is shown in \fref{HT6}. Starting from the top-left corner of the figure, the first graph, $G_{1}$, implements the first $T$ gate, while $G_{2}$, $G_{3}$, and $G_{4}$ implement the first $H$ gate. The sequence is repeated 6 times. Hence, each of the 6 rows in \fref{HT6} implements the quantum operation $HT$. If the adjacency matrix of $G_i$ is denoted $A_i$, then altogether, $U = (HT)^6 = e^{-iA_{24}t_{24}} \cdots e^{-iA_2t_2} e^{-iA_1t_1}$. The total evolution time is $t_1 + t_2 + \dots + t_{24} = 30\pi \approx 94.25$.

Other single-qubit gates can take even more than twenty-four graphs. For example, one way to interpret $U$ is as a rotation by angle
\[ \gamma = 6 \cos^{-1}\left( \frac{\sqrt{2} + 1}{2\sqrt{2}}\right) \]
about the axis
\[ \hat{n} = \left( \frac{1}{\sqrt{5 - 2\sqrt{2}}}, \frac{1 - \sqrt{2}}{\sqrt{5 - 2\sqrt{2}}} , \frac{1}{\sqrt{5 - 2\sqrt{2}}} \right) \]
on the Bloch sphere. Appendix~\ref{appendix:decomposition} reviews how to rewrite single-qubit gates as rotations. If we consider the single-qubit gate that instead rotates by angle $10\gamma$, then $U = (HT)^{60}$, and the number of graphs in the sequence is now 240. This is motivation to find ways to develop shorter dynamic graphs.

\begin{figure} 
\begin{center}
	\includegraphics{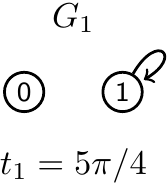} \quad
	\includegraphics{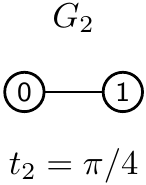} \quad
	\includegraphics{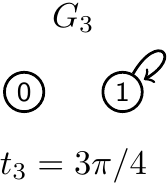} \quad
	\includegraphics{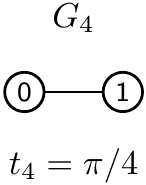}
	\vspace{0.2in}
    
	\includegraphics{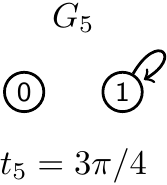} \quad
	\includegraphics{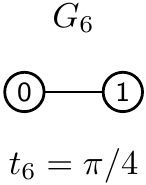} \quad
	\includegraphics{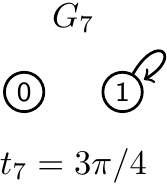} \quad
	\includegraphics{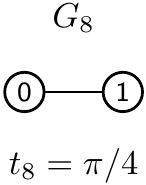}
	\vspace{0.2in}

	\includegraphics{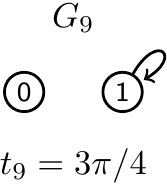} \quad
	\includegraphics{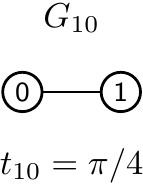} \quad
	\includegraphics{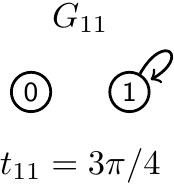} \quad
	\includegraphics{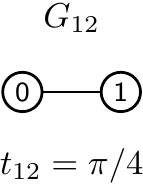}
	\vspace{0.2in}
    
	\includegraphics{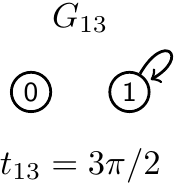} \quad
    
	\caption{\label{HT6_simplified} A simplified dynamic graph that implements $(HT)^{6}$.}
\end{center}
\end{figure}

Several ways to simplify dynamic graphs were explored in \cite{herrman2021}. Using these results, \fref{HT6} can be reduced somewhat. First, $G_1$ and $G_2$ can be combined into a single graph, since they are identical graphs, with an evolution time of $7\pi/4 + 3\pi/2 = 5\pi/4 \pmod{2\pi}$. Next, $G_4$, $G_5$, and $G_6$ can similarly be combined, with an evolution time of $3\pi/2 + 7\pi/4 + 3\pi/2 = 3\pi/4 \pmod{2\pi}$. This same simplification can be applied to $G_8$, $G_9$, and $G_{10}$; $G_{12}$, $G_{13}$, and $G_{14}$; $G_{16}$, $G_{17}$, and $G_{18}$; and $G_{20}$, $G_{21}$, and $G_{22}$. The resulting simplified graph is shown in \fref{HT6_simplified}. This reduces the total number of graphs to thirteen, with a total evolution time of $5\pi/4 + 6(\pi/4) + 5(3\pi/4) + 3\pi/2 = 8\pi \approx 25.13$, which is a 73\% speedup over the original runtime of $30\pi$.

In this paper, we will reduce the number of graphs to just three, with a total evolution time of $367\pi/100 \approx 11.53$, which is an 88\% speedup over the original runtime of $30\pi$. We do this by giving a parameterized dynamic graph of length 3 that can implement any single-qubit quantum gate. This allows us to directly implement any single qubit gate without first decomposing the gate into $H$ and $T$ gates (which make up a universal gate set for single qubits). In some cases, the length of the graph can be reduced to 2 or even 1. In the next section, we will express this. Afterward, in Section III, we will generalize this result to single-qubit quantum gates controlled by any number of qubits, implementing them with dynamic graphs of length at most 3. In Section IV, we will apply our constructions by implementing Draper's quantum addition circuit \cite{draper2000addition} as a dynamic quantum walk. Finally, we conclude in Section V.


\section{Single Qubit Gates}

To begin, note that any single-qubit gate can be written in the \zyz decomposition, i.e., as a rotation about the $z$-axis of the Bloch sphere by some angle $\lambda$, $y$-axis by some angle $\theta$, and $z$-axis by some angle $\phi$, up to a global, irrelevant phase. As a matrix, the \zyz decomposition takes the form
\begin{align}
    U_{\left ( \theta, \phi, \lambda \right)} 
        &= R_{z}(\phi)R_{y}(\theta)R_{z}(\lambda) \nonumber \\
        &= \begin{pmatrix} 
            \cos({\frac{\theta}{2}}) & -e^{i\lambda} \sin({\frac{\theta}{2}}) \\ 
            e^{i\phi} \sin({\frac{\theta}{2}}) & e^{i(\phi + \lambda)} \cos({\frac{\theta}{2}})
        \end{pmatrix}, \label{eq:ZYZ} 
\end{align}
where $R_{n}(\gamma)$ is a rotation on the Bloch sphere about axis $n \in {x,y,z}$ by angle $\gamma$. Appendix~\ref{appendix:decomposition} reviews how to decompose single-qubit gates in this form. Now, if $|\psi\rangle = c_0 |0\rangle + c_1 |1\rangle$ is a general single-qubit state, then the state after applying $U_{(\theta,\phi,\lambda)}$ is
\begin{align}
	U_{\left ( \theta, \phi, \lambda \right)} | \psi \rangle 
		&= \left[c_{0}\cos\left({\frac{\theta}{2}}\right) - e^{i\lambda}c_{1}\sin\left({\frac{\theta}{2}}\right)\right]|0\rangle \label{general_eqn} \\
		&\quad\quad + e^{i\phi}\left[c_{0}\sin\left({\frac{\theta}{2}}\right) + e^{i\lambda}c_{1}\cos\left({\frac{\theta}{2}}\right)\right]|1\rangle. \nonumber
\end{align}

\begin{figure} 
\begin{center}
	\includegraphics{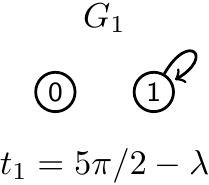} \quad
	\includegraphics{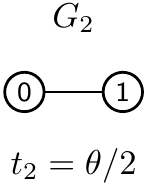} \quad
	\includegraphics{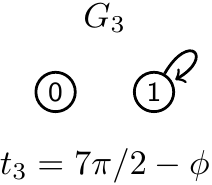}

	\caption{\label{single_length3} A length-3 parameterized dynamic graph that implements a single-qubit operation. Each evolution time $t_i$ should be taken modulo $2\pi$.}
\end{center}
\end{figure}

Next, we show that a continuous-time quantum walk on the length-3 dynamic graph shown in \fref{single_length3} produces \eqref{general_eqn}, thus implementing any single-qubit gate in just three graphs. First, the adjacency matrix of $G_1$ is
\[ A_1 = \begin{pmatrix} 0 & 0 \\ 0 & 1 \end{pmatrix}, \]
and so the time-evolution operator when walking on this graph for time $t_1$ is
\[ e^{-iA_1t_1} = \begin{pmatrix} 1 & 0 \\ 0 & e^{it_1} \end{pmatrix}. \]
If the initial state of the walker is
\[ | \psi_0 \rangle = c_0 | 0 \rangle + c_1 | 1 \rangle = \begin{pmatrix} c_0 \\ c_1 \end{pmatrix}, \]
then after walking on graph $G_1$ for time $t_1$, the state of the walker is
\begin{align*}
	| \psi_1 \rangle
		&= e^{-iA_1t_1} | \psi_0 \rangle = \begin{pmatrix} 1 & 0 \\ 0 & e^{it_1} \end{pmatrix} \begin{pmatrix} c_0 \\ c_1 \end{pmatrix} \\
		&= \begin{pmatrix} c_0 \\ e^{-it_1} c_1 \end{pmatrix} = c_0 | 0 \rangle + e^{-it_1} c_1 | 1 \rangle.
\end{align*}
In our graph in \fref{single_length3}, $t_{1} = 5\pi/2 - \lambda \pmod{2\pi}$, so the state of the walker after the first graph is
\begin{align*}
	| \psi_1 \rangle
		&= c_{0} |0\rangle + e^{-i(5\pi/2 - \lambda)} c_{1} |1\rangle \\
		&= c_{0} |0\rangle + e^{-i\pi/2} e^{-i(2\pi - \lambda)} c_{1} |1\rangle \\
		&= c_{0} |0\rangle - i e^{i\lambda} c_{1} |1\rangle.
\end{align*}
Next, the adjacency matrix of $G_2$ and the time-evolution under this graph for time $t_2$ are
\[ A_2 = \begin{pmatrix} 0 & 1 \\ 1 & 0 \end{pmatrix}, \quad e^{-iA_2t_2} = \begin{pmatrix} \cos(t_2) & -i \sin(t_2) \\ -i \sin(t_2) & \cos(t_2) \end{pmatrix}. \]
So, if we evolve by $G_{2}$ for time $t_{2} = \theta/2$, the state of the walker becomes
\begin{align*}
	| \psi_2 \rangle
		&= e^{-iA_2t_2} | \psi_1 \rangle \\
		&= \left[c_{0}\cos{\left(\frac{\theta}{2}\right)} - e^{i\lambda}c_{1}\sin{\left(\frac{\theta}{2}\right)}\right]|0\rangle \\  
		&\quad + \left[-ie^{i\lambda}c_{1}\cos{\left(\frac{\theta}{2}\right)} - ic_{0}\sin{\left(\frac{\theta}{2}\right)}\right]|1\rangle.
\end{align*}
Finally, $G_3$ is the same as $G_1$, except the evolution time is $t_3 = 7\pi/2 - \phi \pmod{2\pi}$. After evolving on $G_3$, the final state of the walker is
\begin{align*}
	| \psi_3 \rangle
		&= e^{-iA_3t_3} | \psi_2 \rangle \\
		&= \left[c_{0}\cos\left({\frac{\theta}{2}}\right) - e^{i\lambda}c_{1}\sin\left({\frac{\theta}{2}}\right)\right]|0\rangle \\ 
&\quad + e^{-i(7\pi/2 - \phi)} \left[-ie^{i\lambda}c_{1}\cos{\left(\frac{\theta}{2}\right)} - ic_{0}\sin{\left(\frac{\theta}{2}\right)}\right]|1\rangle.
\end{align*}
Since
\[ e^{-i(7\pi/2 - \phi)} = e^{-i3\pi/2} e^{-i(2\pi - \phi)} = i e^{i\phi}, \]
the final state is
\begin{align}
	| \psi_3 \rangle
		&= \left[c_{0}\cos\left({\frac{\theta}{2}}\right) - e^{i\lambda}c_{1}\sin\left({\frac{\theta}{2}}\right)\right]|0\rangle \label{general_eqn_2} \\ 
		&\quad + e^{i\phi}\left[c_{0}\sin\left({\frac{\theta}{2}}\right) + e^{i\lambda}c_{1}\cos\left({\frac{\theta}{2}}\right)\right]|1\rangle. \nonumber
\end{align}
This is exactly the same as \eqref{general_eqn}. Since every single-qubit operation can be written in form of $U_{(\theta, \phi, \lambda)}$, up to a global phase, the length-3 dynamic graph in \fref{single_length3} can implement any single-qubit operation with the appropriate choice of $\theta$, $\phi$, and $\lambda$. Moreover, since each of these angles are in the interval $[0,2\pi)$, and the evolution times in \fref{single_length3} are taken modulo $2\pi$, we have $t_1 < 2\pi$, $t_2 < \pi$, and $t_3 < 2\pi$. So, the total time needed to implement a single-qubit gate using our construction is less than $2\pi + \pi + 2\pi = 5\pi$. Finally, if one wishes to apply a physically irrelevant global phase of $e^{i\alpha}$, one can walk on a fourth graph with a self-loop on each vertex for time $t_4 = 2\pi - \alpha \pmod{2\pi}$.

\begin{figure} 
\begin{center}
	\includegraphics{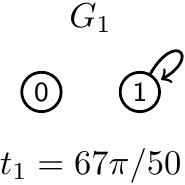} \quad
	\includegraphics{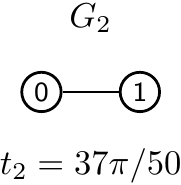} \quad
	\includegraphics{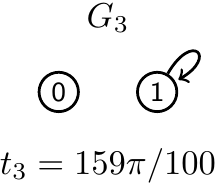}

	\caption{\label{HT6_length3} A length-3 dynamic graph that implements $(HT)^{6}$.}
\end{center}
\end{figure}

Let us apply this to the example from the introduction, where we considered the single-qubit gate $U = (HT)^6$. Rewriting it in the \zyz decomposition up to a global phase, we get (see Appendix~\ref{appendix:decomposition})
\[ (HT)^{6} = U_{(37\pi/25, 191\pi/100, 29\pi/25)}. \]
If we then put the values of these parameters $(\theta, \phi, \lambda)$ into \fref{single_length3}, we get
\begin{align*}
    t_1 &= \frac{5\pi}{2} - \lambda = \frac{5\pi}{2} - \frac{29\pi}{25} = \frac{67\pi}{50}, \\
    t_2 &= \frac{\theta}{2} = \frac{37\pi}{2\cdot25} = \frac{37\pi}{50}, \\
    t_3 &= \frac{7\pi}{2} - \phi = \frac{7\pi}{2} - \frac{191\pi}{100} = \frac{159\pi}{100}.
\end{align*}
With these times, the graph that implements $(HT)^{6}$ is shown in \fref{HT6_length3}, and its total evolution time is $367\pi/100 \approx 11.53$. As described in the introduction, this is a significant improvement over previous results, where \fref{HT6} used 24 graphs and a total evolution time of 94.25, and \fref{HT6_simplified} used 13 graphs with a total evolution time of $25.13$.

Our improvement can be even more dramatic for other single-qubit gates. For example, if we wish to perform $(HT)^{60}$, the maximum length of our implementation is 3 graphs. This is unlike the previous results in \cite{wong2019isolated} and \cite{herrman2021} that use a length-240 dynamic graph and a length-121 dynamic graph, respectively, to implement the same quantum operation.

\begin{figure} 
\begin{center}
	\subfloat[] {
		\label{single_length3_lambda} 
		\includegraphics{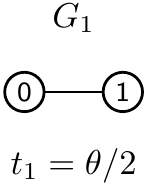} \quad
		\includegraphics{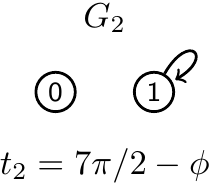}
	}

	\subfloat[] {
		\label{single_length3_phi} 
		\includegraphics{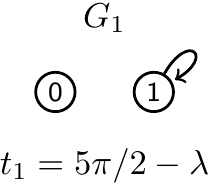} \quad
		\includegraphics{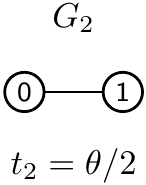}
	}

	\subfloat[] {
		\label{single_length3_theta}
		\includegraphics{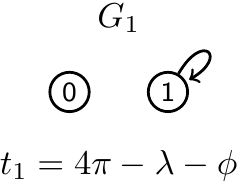}
	}

	\caption{\label{single_length3_simplified}The dynamic quantum walk from \fref{single_length3} that implements the single-qubit gate $U_{(\theta, \phi, \lambda)}$, simplified when (a) $\lambda = \pi/2$, (b) $\phi = 3\pi/2$, and (c) $\theta = 0$. Each evolution time $t_i$ should be taken modulo $2\pi$.}
\end{center}
\end{figure}

In some cases, our length-3 dynamic graph can be reduced further, depending on the parameters $\theta$, $\phi$, and $\lambda$. For each of the graphs in \fref{single_length3}, an evolution for time $2\pi$ is equivalent to an identity operation (no evolution) \cite{wong2019isolated}. With this, we can reduce the length of the dynamic graph when:  
\begin{enumerate}
	\item 	$\lambda = \pi/2$. When this is true, we can ignore $G_{1}$. This is because for $\lambda = \pi/2$, the evolution time for the first graph becomes $t_1 = 5\pi/2 - \pi/2 = 2\pi$. Since there is no evolution when $t_1 = 2\pi$, we can then ignore the first graph, $G_{1}$. Hence, the dynamic graph in \fref{single_length3} reduces to \fref{single_length3_lambda}. An example of such an operation is the Pauli $Y$ gate, where $Y = U_{(\pi, \pi/2, \pi/2)}$.

	\item 	$\phi = 3\pi/2$. Following the same logic as above, we can then ignore the last graph, $G_{3}$, reducing the length of the dynamic graph in \fref{single_length3} to two as in \fref{single_length3_phi}. 
    
	\item 	$\theta = 0$. When this is true, we can ignore the second graph, $G_{2}$, and the total length will reduce to just 1. This is because the first and the last static graphs are the same graph, so we can combine them into one graph with evolution time $t_1 + t_3 = 5\pi/2 - \lambda + 7\pi/2 - \phi = 6\pi - \lambda - \phi = 4\pi -\lambda - \phi \pmod{2\pi}$. For this case, the dynamic graph is reduced to \fref{single_length3_theta}.
\end{enumerate}


\section{Controlled Gates}

\begin{figure} 
\begin{center}
	\subfloat[] {
	    \label{ctrl_op_dgraph}
		\includegraphics{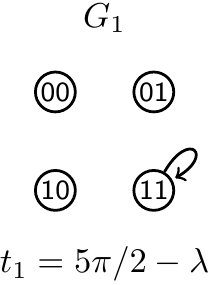} \quad
		\includegraphics{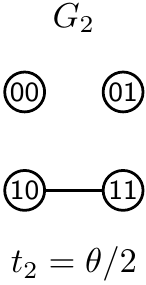} \quad
		\includegraphics{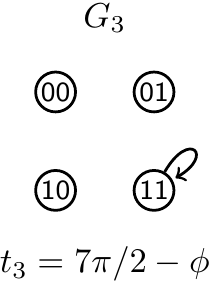}
	}

	\subfloat[] {
	    \label{ctrl_ctrl_op_dgraph}
		\includegraphics{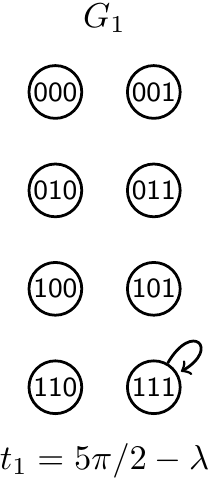} \quad
		\includegraphics{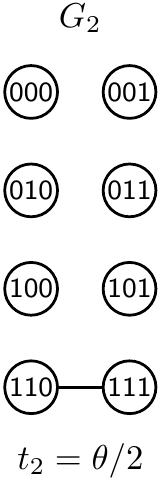} \quad
		\includegraphics{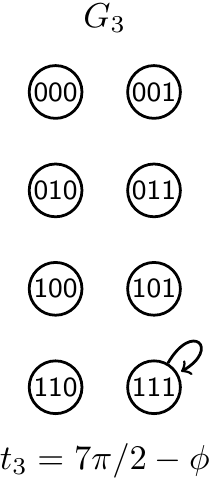}
	}
        
	\caption{A dynamic graph on (a) two qubits that applies the unitary $U_{(\theta,\phi,\lambda)}$ on the right qubit, controlled by the left qubit, and (b) three qubits that applies the unitary $U_{(\theta,\phi,\lambda)}$ on the right qubit, controlled by the left and middle qubits. Each evolution time $t_i$ should be taken modulo $2\pi$.}
\end{center}
\end{figure}

In this section, we generalize our previous length-3 dynamic graph for single-qubit gates (i.e., \fref{single_length3}) so that the single-qubit gates can be controlled by any number of qubits. The simplest case is when there is just one control qubit, plus the target qubit. We denote this by $C(U)$, and it acts on the four basis states by
\begin{align*}
    C(U) | 00 \rangle &= | 00 \rangle, \\
    C(U) | 01 \rangle &= | 01 \rangle, \\
    C(U) | 10 \rangle &= (I \otimes U) | 10 \rangle, \\
    C(U) | 11 \rangle &= (I \otimes U) | 11 \rangle.
\end{align*}
That is, $U$ is applied to the second qubit when the first qubit is 1. Generalizing \fref{single_length3}, the dynamic graph for $C(U)$ is shown in \fref{ctrl_op_dgraph}. It has four vertices for the four computational basis states. Nothing happens to vertices 00 and 01 because $C(U)$ does nothing to these basis elements. When the left qubit is 1, it applies $U_{(\theta,\phi,\lambda)}$, so vertices $10$ and $11$ are the same as what we had in \fref{single_length3}.

Similarly, if we have the controlled-controlled-unitary $C^2(U)$, our generalization yields the length-3 dynamic graph in \fref{ctrl_ctrl_op_dgraph}. Now, the vertices $110$ and $111$, i.e., where the left two control qubits are both 1, are evolving. We can generalize further to any arbitrary number of qubits, say $C^k(U)$. The graph would have $2^{k+1}$ vertices. The vertices $11\dots10$ and $11\dots11$ would evolve while the rest would do nothing. Finally, note in all of these generalizations, we can shorten the graphs as in \fref{single_length3_simplified}.


\section{Quantum Addition Simulation}

In this section, we will verify our results by simulating a quantum addition circuit as a dynamic quantum walk. In \cite{herrman2019continuous}, the quantum version of the classical ripple quantum adder from \cite{Vedral1996} was simulated. This ripple-carry adder only uses CNOT and Toffoli gates, and since these gates are simply Pauli $X$ gates controlled by one or two other qubits, respectively, they can be implemented using our dynamic graphs of length at most three, but this would result in essentially the same implementation as \cite{herrman2019continuous}. Furthermore, the quantum ripple-carry adder requires an extra quantum register for the carry bits. So instead, we will simulate Draper's quantum adder \cite{draper2000addition}, which does not require an extra register for the carry bits. It is based on the quantum Fourier transform and only uses single-qubit gates, and single-qubit gates controlled by a second qubit, and all of these gates can be naturally implemented using our dynamic graphs of length at most three.

To add two length-$n$ binary numbers $a = a_n \dots a_1$ and $b = b_n \dots b_1$, Draper's adder takes
\[ | b \rangle | a \rangle \to | b \rangle | a + b \rangle. \]
Or in terms of the bits, the adder takes
\[ | b_n \dots b_1 \rangle | a_n \dots a_1 \rangle \to | b_n \dots b_1 \rangle | (a+b)_n \dots (a+b)_1 \rangle. \]
Furthermore, since the adder is quantum, it can add numbers in superposition. For example, with $n = 3$, say $a$ is a uniform superposition of 2 and 3, and $b$ is a uniform superposition of 1 and 3, i.e.,
\begin{align*}
	&| a \rangle = \frac{1}{\sqrt{2}} \left( | 2 \rangle + | 3 \rangle \right) = \frac{1}{\sqrt{2}}\left(|010\rangle + |011\rangle \right), \\
	&| b \rangle = \frac{1}{\sqrt{2}} \left( | 1 \rangle + | 3 \rangle \right) = \frac{1}{\sqrt{2}}\left( |001\rangle + |011\rangle \right).
\end{align*}
Then, the adder takes as input
\begin{align}
	| b \rangle | a \rangle 
		&= \frac{1}{\sqrt{2}} \left( | 1 \rangle + | 3 \rangle \right) \frac{1}{\sqrt{2}} \left( | 2 \rangle + | 3 \rangle \right) \nonumber \\
		&= \frac{1}{2} \left( | 1,2 \rangle + | 1,3 \rangle + | 3, 2 \rangle + | 3,3 \rangle \right) \nonumber \\
		&= \frac{1}{2} ( | 001010 \rangle + | 001011 \rangle \label{eq:sim_initial_together} \\
		&\quad\quad + | 011010 \rangle + | 011011 \rangle ). \nonumber
\end{align}
With this input, the sum would be a uniform superposition of $2+1=3$, $3+1=4$, $2+3=5$, and $3+3=6$, i.e., the final state after the adder would be
\begin{align}
	| b \rangle | (a+b) \rangle 
		&= \frac{1}{2} \left( | 1,3 \rangle + | 1, 4 \rangle + | 3, 5 \rangle + | 3, 6 \rangle \right) \nonumber \\
		&= \frac{1}{2} ( | 001011 \rangle + | 001100 \rangle \label{addition_result} \\
		&\quad\quad + | 011101 \rangle + | 011110 \rangle ). \nonumber
\end{align}

\begin{figure*}
	\includegraphics{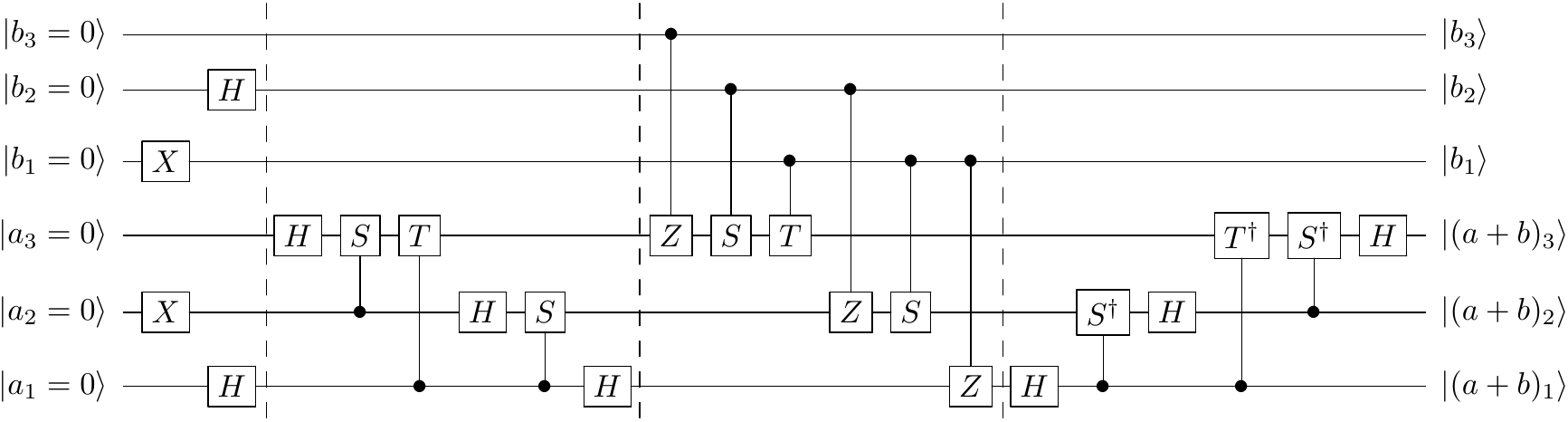}
	\caption{\label{q_addition_circuit}A 20-layer quantum addition circuit of two 3-qubit integers. Vertical dashed lines separate the circuit into four sections. The first section sets the bitstrings to be added in superposition. The second section is a quantum Fourier transform. The third section is Draper's addition. The fourth section is the inverse Fourier transform.} 
\end{figure*}

Draper's circuit for performing this computation is shown in \fref{q_addition_circuit}. The quantum circuit has six qubits. The top three qubits represent the three qubits of input $|b\rangle$, and the last three qubits represents the three qubits of input $|a\rangle$. This means the qubits are arranged as $|b_{3}b_{2}b_{1}a_{3}a_{2}a_{1}\rangle$, with $|b_{3}\rangle$ being the topmost qubit in the circuit (we also refer to it as the first qubit), while $|a_{1}\rangle$ is the qubit at the bottom of the circuit (also referenced as the last or sixth qubit). The vertical dashed lines demarcate the sections of the circuit. The first section creates the input state \eqref{eq:sim_initial_together}. The second, third, and fourth sections make up Draper's quantum addition circuit \cite{draper2000addition}. As described in \cite{draper2000addition}, the second section computes the quantum Fourier transform (QFT) of $|a\rangle$, denoted $|\phi(a)\rangle$. The third section adds the bits, transforming $|\phi(a)\rangle$ to $|\phi(a+b)\rangle$. The last section performs the inverse QFT, which transforms $|\phi(a+b)\rangle$ to $|a+b\rangle$, which results in \eqref{addition_result}. The complete evolution of the circuit is shown in Appendix~\ref{appendix:addition}, and \eqref{first_eqn} shows the initial state of the system, \eqref{barrier_1} gives the state after the first section of the circuit, \eqref{barrier_2} after the second section, \eqref{barrier_3} after the third section, and \eqref{final_eqn} the final state. In the rest of this section, we will implement \eqref{q_addition_circuit} using a dynamic quantum walk and present a simulation of the walk.

\begin{figure*}
	\includegraphics{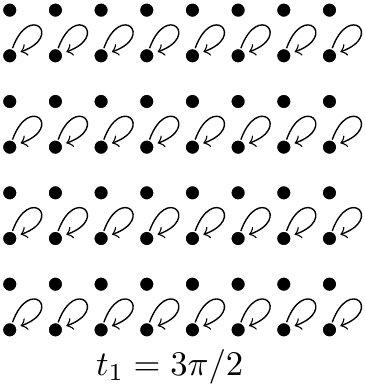} \quad
	\includegraphics{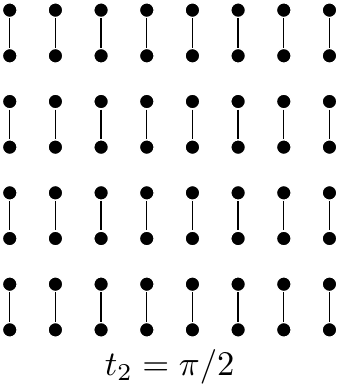} \quad
	\includegraphics{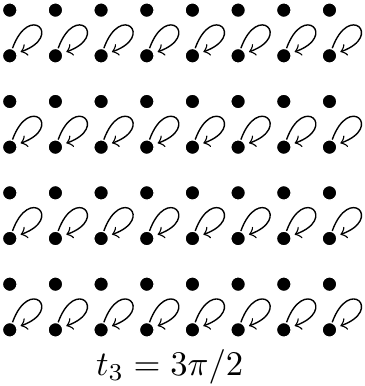} \quad
	\includegraphics{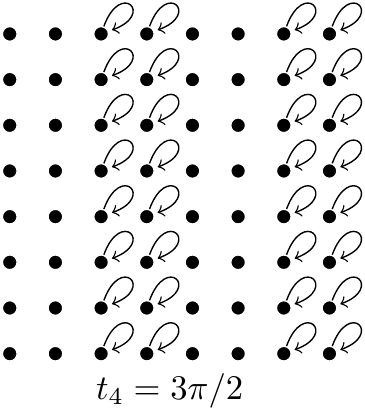}
	\vspace{0.1in}

	\includegraphics{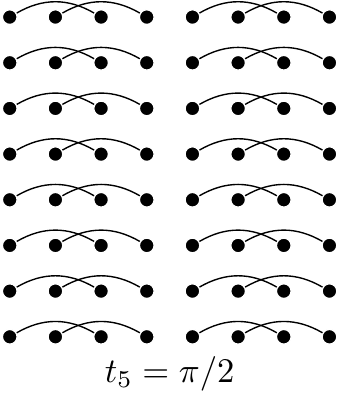} \quad
	\includegraphics{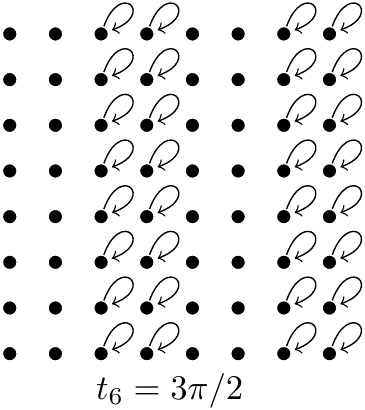} \quad
	\includegraphics{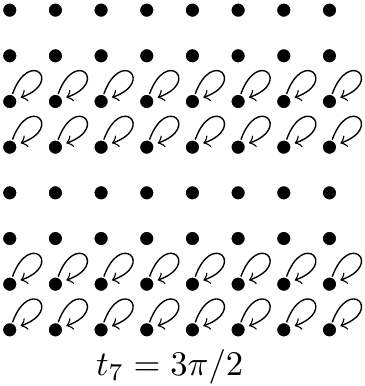} \quad
	\includegraphics{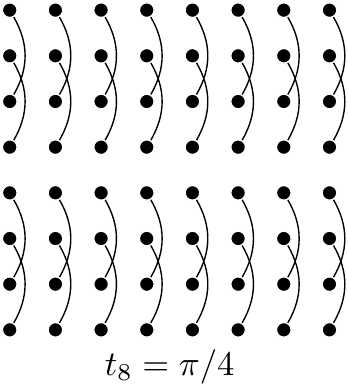}
	\vspace{0.1in}

	\includegraphics{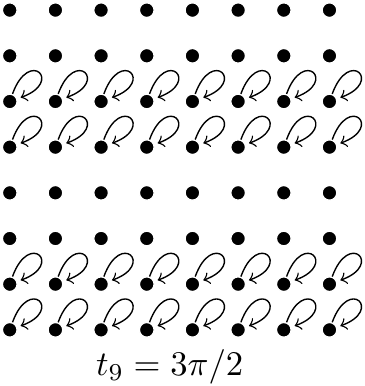} \quad
	\includegraphics{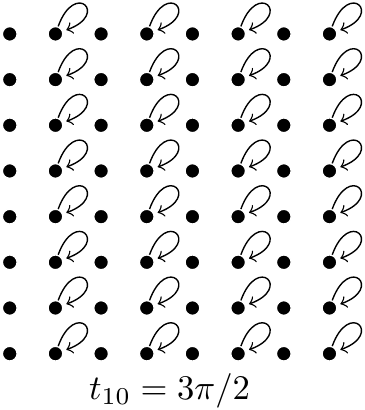} \quad
	\includegraphics{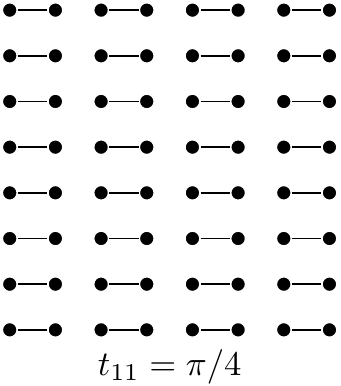} \quad
	\includegraphics{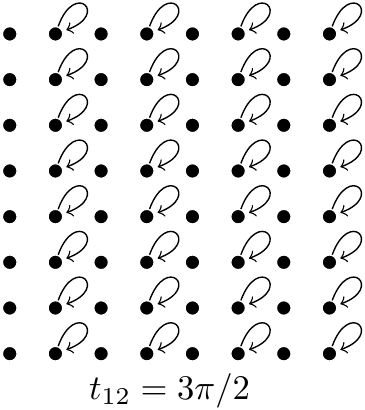}
        \caption{\label{superposition} A dynamic graph on which a quantum walk implements the first section of \fref{q_addition_circuit}.}
\end{figure*}

The dynamic graph that implements \fref{q_addition_circuit} is 42 graphs long, so we will present it by section. The first section of \fref{q_addition_circuit} is implemented by the dynamic graph in \fref{superposition}. Each graph has 64 vertices, denoted by solid black dots, and they are ordered left-to-right, top-to-bottom, from $|000000\rangle$ to $|111111\rangle$. In \fref{superposition}, graphs $G_1$, $G_2$, and $G_3$ implement $X$ on the third qubit, while graphs $G_4$, $G_5$, and $G_6$ implement $X$ on the fifth qubit. Graphs $G_7$, $G_8$, and $G_9$ implement $H$ on the second qubit, and the last three graphs implement $H$ on the last qubit. 

\begin{figure*}
\begin{center}
	\includegraphics{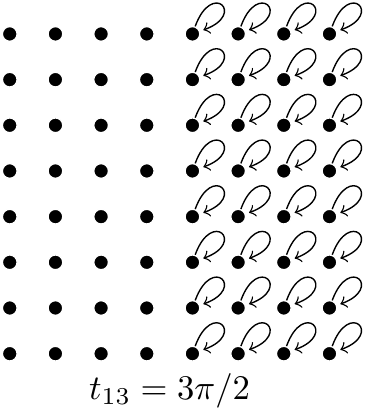} \quad
	\includegraphics{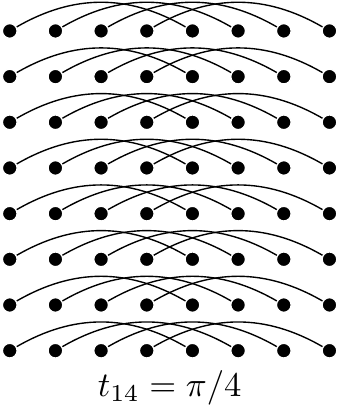} \quad
	\includegraphics{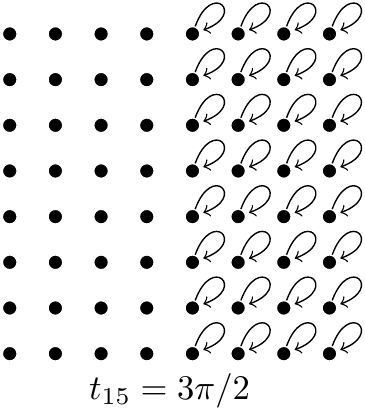} \quad
	\includegraphics{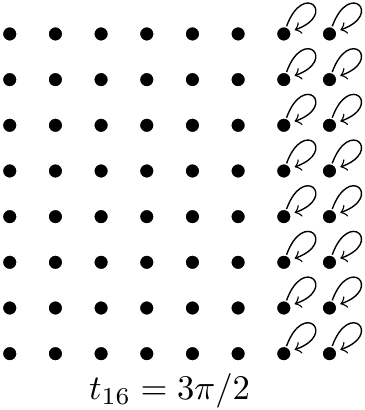}
	\vspace{0.1in}

	\includegraphics{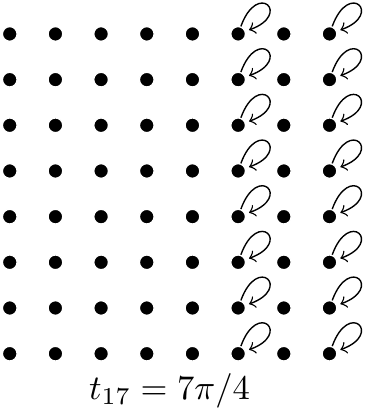} \quad
	\includegraphics{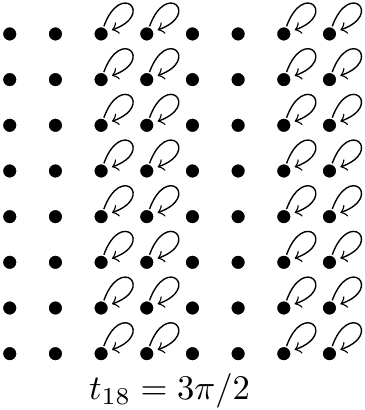} \quad
	\includegraphics{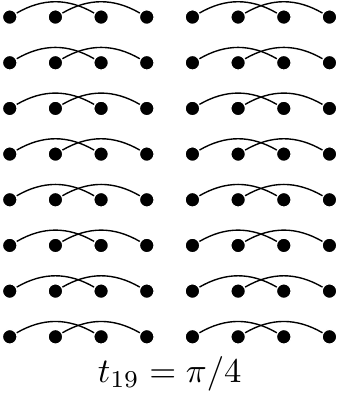} \quad
	\includegraphics{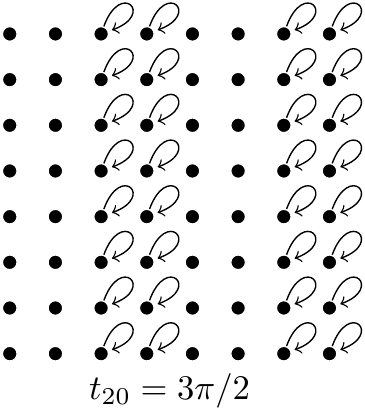}
	\vspace{0.1in}

	\includegraphics{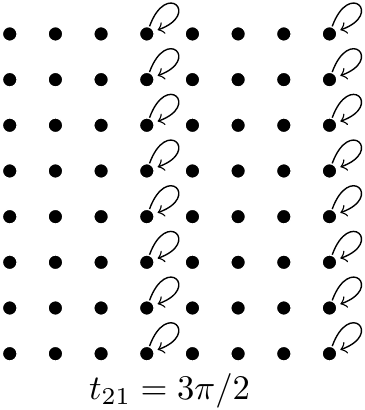} \quad
	\includegraphics{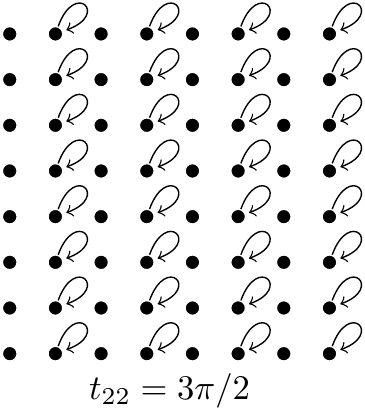} \quad
	\includegraphics{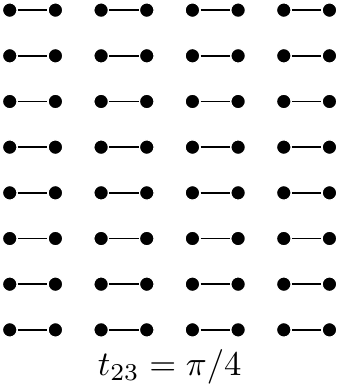} \quad
	\includegraphics{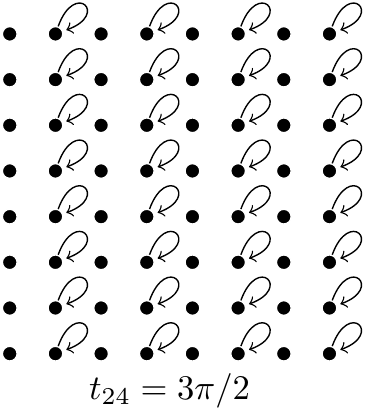}

	\caption{\label{GraphsforQFT} A dynamic graph on which a quantum walk implements the second section of \fref{q_addition_circuit}.}
\end{center}
\end{figure*}

Next, the dynamic graph in \fref{GraphsforQFT} implements the second section of \fref{q_addition_circuit}, which computes $|\phi(a)\rangle$. Graphs $G_{13}$, $G_{14}$, and $G_{15}$ implement $H$ on the fourth qubit. Graph $G_{16}$ implements $C(S_{5,4})$, where the subscripts denote that the fifth qubit is the control and the fourth qubit is the target. $G_{17}$ implements $C(T_{6,4})$, while $G_{18}$, $G_{19}$, and $G_{20}$ implement $H$ on the fifth qubit. $G_{21}$ implements $C(S_{6,5})$, and $G_{22}$, $G_{23}$, and $G_{24}$ implement $H$ on the sixth qubit.

\begin{figure*}
	\includegraphics{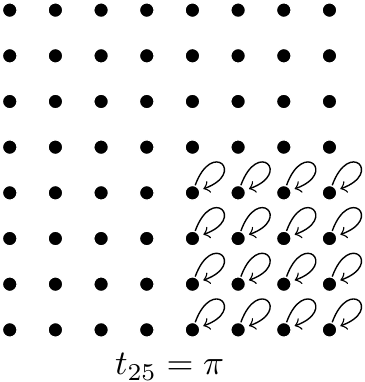} \quad
	\includegraphics{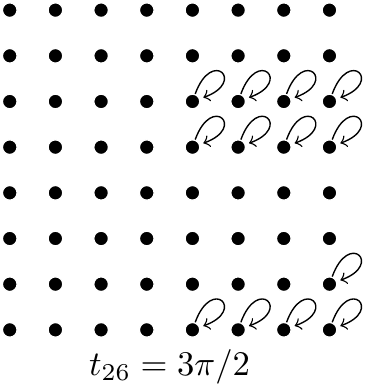} \quad
	\includegraphics{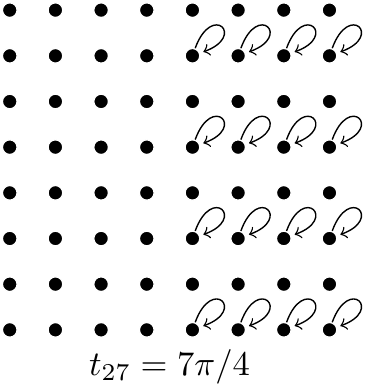} \quad
	\includegraphics{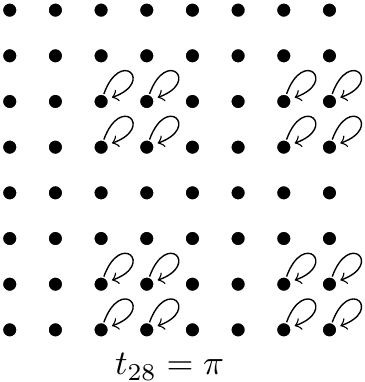}
	\vspace{0.1in}

	\includegraphics{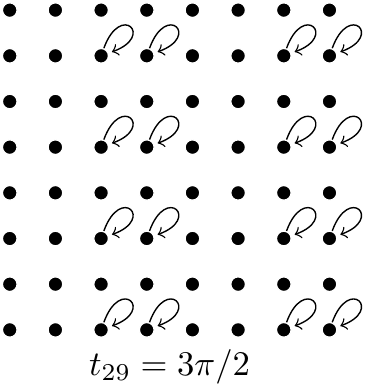} \quad
	\includegraphics{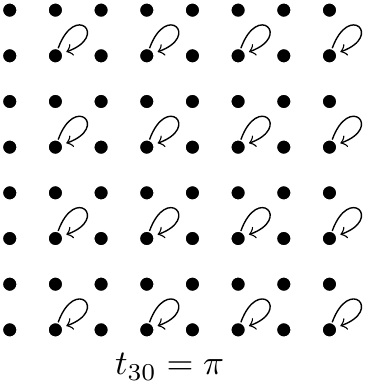} \quad

	\caption{\label{GraphsforDraper} A dynamic graph on which a quantum walk implements the third section of \fref{q_addition_circuit}.}

\end{figure*}

Similarly, \fref{GraphsforDraper} implements the third section of \fref{q_addition_circuit}, which computes $|\phi(a+b)\rangle$. $G_{25}$ implements $C(Z_{1,4})$, $G_{26}$ implements $C(S_{2,4})$, $G_{27}$ implements $C(T_{3,4})$, $G_{28}$ implements $C(Z_{2,5})$, $G_{29}$ implements $C(S_{3,5})$, and $G_{30}$ implements $C(Z_{3,6})$.

\begin{figure*}
	\includegraphics{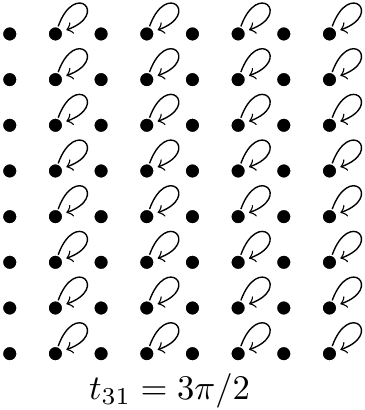} \quad
	\includegraphics{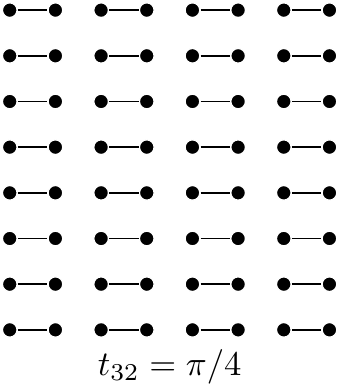} \quad
	\includegraphics{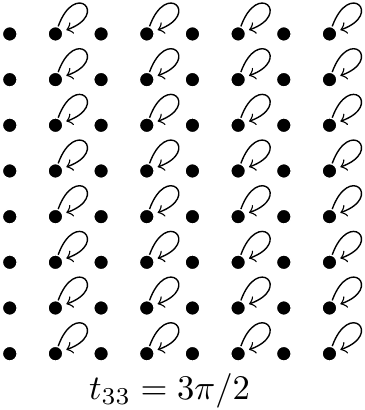} \quad
	\includegraphics{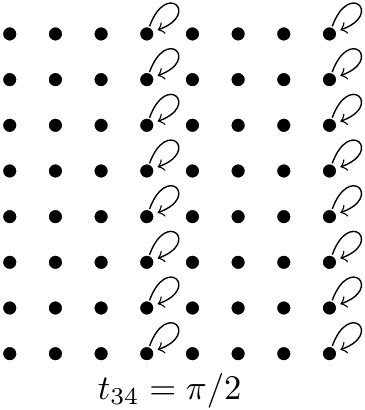}
	\vspace{0.1in}

	\includegraphics{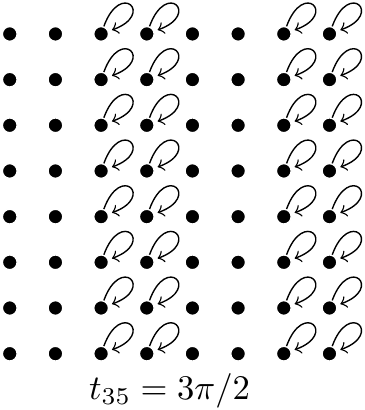} \quad
	\includegraphics{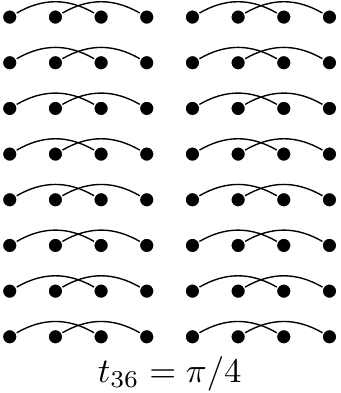} \quad
	\includegraphics{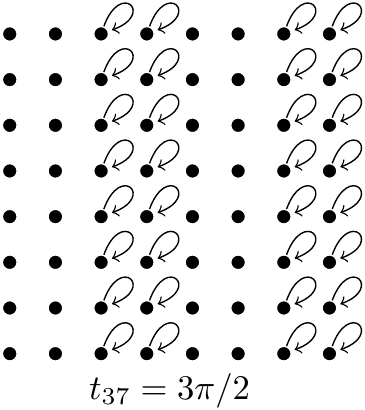} \quad
	\includegraphics{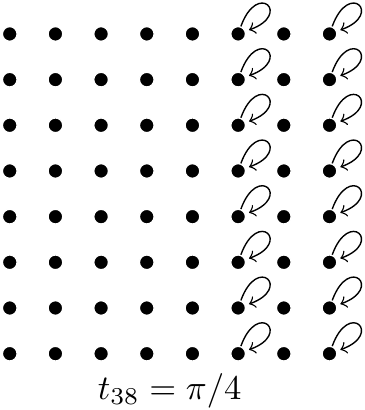}
	\vspace{0.1in}

	\includegraphics{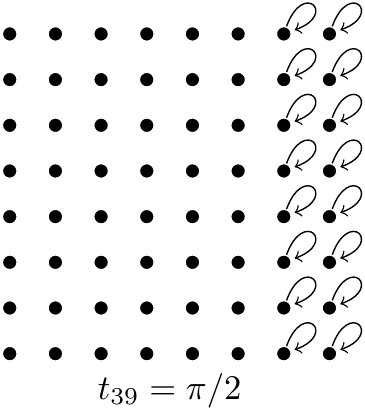} \quad
	\includegraphics{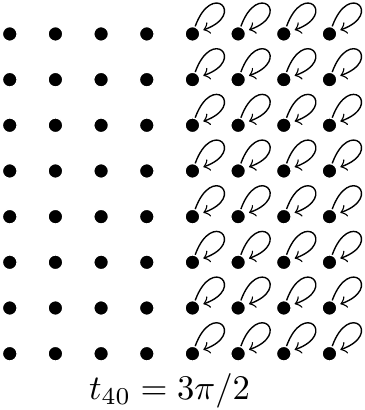} \quad
	\includegraphics{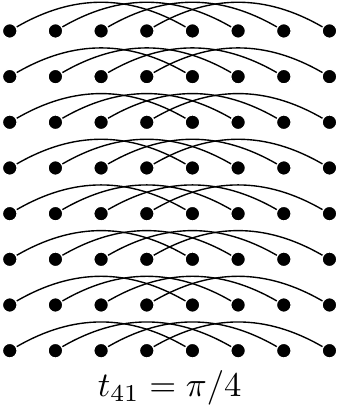} \quad
	\includegraphics{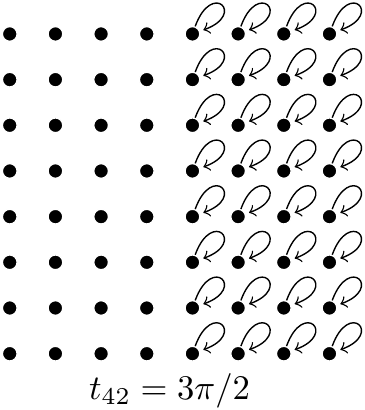}

        \caption{\label{GraphsforinverseQFT} A dynamic graph on which a quantum walk implements the fourth section of \fref{q_addition_circuit}.}
\end{figure*}

Lastly, \fref{GraphsforinverseQFT} implements the fourth section of \fref{q_addition_circuit}, which computes the inverse Fourier transform of $|\phi(a+b)\rangle$. The first three graphs, $G_{31}$, $G_{32}$, and $G_{33}$, implement $H$ on the last qubit, and $G_{34}$ implements $C(S_{6,5}^{\dagger})$. Graphs $G_{35}$, $G_{36}$, and $G_{37}$ implement $H$ on the fifth qubit, $G_{38}$ implements $C(T_{6,4}^{\dagger})$, $G_{39}$ implements $C(S_{5,4}^{\dagger})$, and the last three graphs $G_{40}$, $G_{41}$, and $G_{42}$ implement $H$ on the fourth qubit. The total evolution time across all forty-two graphs is $187\pi/4 \approx 146.87$.

\begin{figure*} 
	\subfloat[]{
		\label{simulation_1}
		\includegraphics{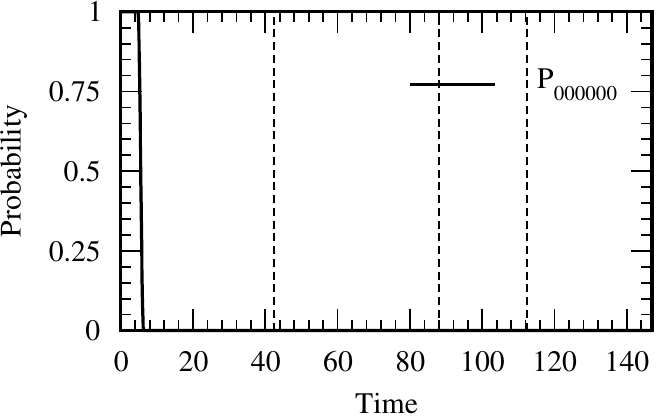}
	} \quad
	\subfloat[]{
		\label{simulation_2}
		\includegraphics{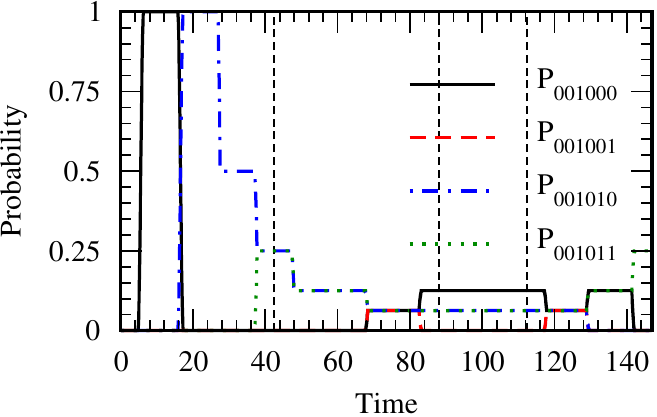}
	}

	\subfloat[]{
		\label{simulation_3}
		\includegraphics{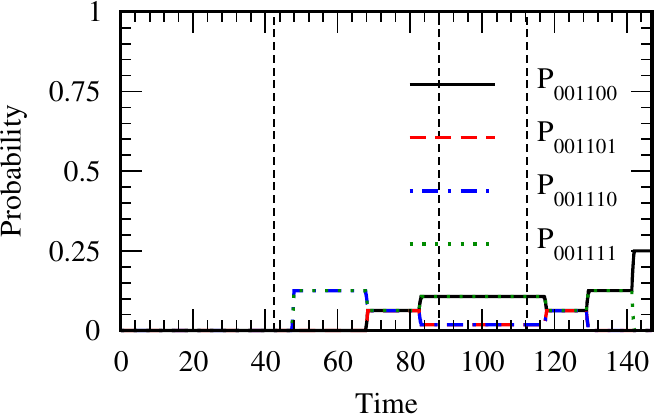}
        } \quad
	\subfloat[]{
		\label{simulation_4}
		\includegraphics{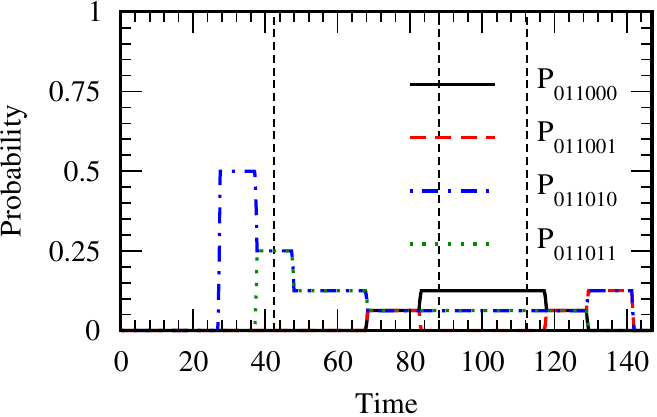}

	}

	\subfloat[]{
		\label{simulation_5}
		\includegraphics{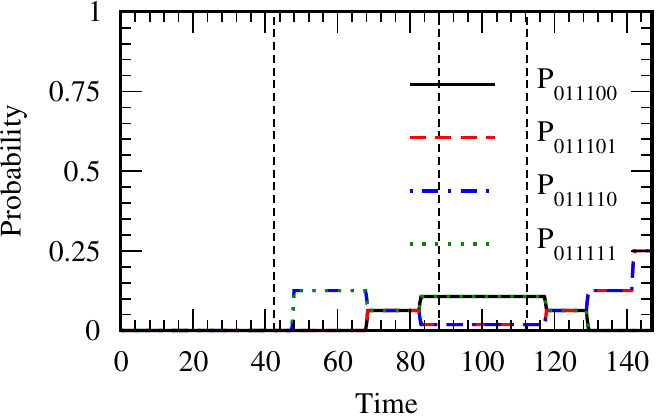}
        }
	\caption{\label{simulation} Time evolution of the vertices $|000000\rangle$ through $|111111\rangle$ due to a quantum walk on dynamic graphs in \fref{superposition} through  \fref{GraphsforinverseQFT}. The vertical dashed lines at times $t = 27\pi/2$, $28\pi$, and $143\pi/4$ mark the different sections of the circuit \fref{q_addition_circuit}.}
\end{figure*}

Now, for the actual simulation of the circuit using a quantum walk on on each of the graphs, \fref{simulation} shows the evolution of each vertex from time $t=0$ to time $t = 187\pi/4 \approx 146.87$. The vertical dashed lines in each plot mark the barriers separating the sections of the circuit in \fref{q_addition_circuit}. The first dashed line is at $t = 27\pi/2 \approx 42.41$, the second dashed line is at $t=28\pi \approx 87.96$, and the third dashed line is at $t=143\pi/4 \approx 112.31$. We have only included the vertices that have nonzero amplitudes and probabilities. All other vertices that are not included have zero amplitudes from time $t=0$ to time $t=187\pi/4$. At the start of the simulation, at time $t=0$, the solid black line in \fref{simulation_1} shows that the probability is 1 at vertex $|000000\rangle$. This agrees with \fref{q_addition_circuit} and \eqref{first_eqn}, where the qubits all start in the $| 0 \rangle$ state.

At the first vertical dashed line (at $t = 27\pi/2$), the first section of the circuit has been applied, and there are four vertices with nonzero probability. They are $|001010\rangle$ and $|001011\rangle$, represented by the dot-dashed blue lines and the dotted green lines, respectively, in \fref{simulation_2}, and $|011010\rangle$ and $|011011\rangle$, also represented by the dot-dashed blue and the dotted green lines, respectively, in \fref{simulation_4}. Each of these vertices has a probability of 1/4 = 0.25, which is consistent with \eqref{eq:sim_initial_together} and \eqref{barrier_1}.

At the second vertical dashed line (at $t = 28\pi$), the second section of the circuit has been applied, and there are fourteen vertices with nonzero amplitudes. Four of the vertices each have a probability of $1/16 = 0.0625$, and they include $|001010\rangle$ and $|001011\rangle$, represented by the dot-dashed blue and dotted green lines, respectively, in \fref{simulation_2}, respectively, as well as $|011010\rangle$ and $|011011\rangle$, also represented by the dot-dashed blue and dotted green lines, respectively, in \fref{simulation_4}. Another four vertices each have a probability of $(2+\sqrt{2})/32 \approx 0.10$, and they include $|001100\rangle$ and $|001111\rangle$, represented by the solid black and dotted green lines, respectively, in figures \fref{simulation_3}, as well as $|011100\rangle$ and $|011111\rangle$, also represented by the solid black and dotted green lines, respectively, in \fref{simulation_5}. Next, another four vertices each have a probability of $(2-\sqrt{2})/32 \approx 0.018$, and they include $|001101\rangle$ and $|001110\rangle$, represented by the dashed red and dot-dashed blue lines, respectively, in \fref{simulation_3}, as well as $|011101\rangle$ and $|011110\rangle$, also represented by the dashed red and dot-dashed blue lines, respectively, in \fref{simulation_5}. Finally, there are two vertices with probability $1/8 = 0.125$, and they are  $|001000\rangle$ and $|011000\rangle$, represented by solid black lines in \fref{simulation_2} and \fref{simulation_4}, respectively. These probabilities are consistent with \eqref{barrier_2}.

These vertices have the same probabilites through the third vertical line at $t = 143\pi/4$. This is because all of the operations in the third section are controlled-rotations around the $z$-axis of the Bloch sphere, so only the phases of the amplitudes change, not their magnitudes. This does not affect the probability distribution, and it is consistent with \eqref{barrier_3}.

Finally, at time $t = 187\pi/4$, which marks the end of the circuit, the only vertices with non-zero probability are $|001011\rangle$ (represented by the dotted green line in \fref{simulation_2}), $|001100\rangle$ (represented by the solid black line in \fref{simulation_3}), $|011101\rangle$ (represented by the dashed red line in \fref{simulation_5}), and $|011110\rangle$ (represented by the dot-dashed blue line in \fref{simulation_5}). Each of these vertices has a probability of $1/4 = 0.25$, in agreement with the expected result in \eqref{addition_result} and \eqref{final_eqn}. This shows that the dynamic quantum walk correctly simulates the quantum addition circuit.  


\section{Conclusion}

A quantum walk evolves on a graph by Schr\"odinger's equation with an appropriate Hamiltonian. With the Hamiltonian equal to the adjacency matrix, quantum walks on dynamic graphs can implement a universal set of quantum gates in as few vertices as possible because each basis state requires just one vertex. With this result, we can implement any arbitrary quantum gate to any desired precision. However, this implementation may be long. In this paper, we have addressed this for single-qubit gates by developing a parameterized dynamic graph on which a continuous-time quantum walk can implement any single-qubit quantum gate with at most three graphs. So, instead of decomposing a single-qubit operation to gates found in a universal set of quantum gates, we implement the single-qubit operation directly. We also extended this result to implement any single-qubit gate controlled by any number of qubits. Finally, we verified our construction by simulating Draper's quantum addition circuit, which is based on the quantum Fourier transform.

Regarding possible physical implementations, Herrman and Humble \cite{herrman2019continuous} described in some length how coupled waveguides might be used to implement dynamic quantum walks, including how the couplings between the waveguides could be controlled. They also suggested that the tunability of the M{\o}lmer-S{\o}rensen gate could be used in systems of trapped ions. We refer readers to \cite{herrman2019continuous} for more detail.


\appendix
\section{\label{appendix:decomposition}Decomposing Single-Qubit Unitaries}

In this appendix, we show how to decompose a single-qubit gate in two different ways. First, as a rotation by angle $\gamma$ about axis $\hat{n} = (n_x,n_y,n_z)$, and second, in the \zyz decomposition.

First, a rotation by angle $\gamma$ about the axis $\hat{n}$ on the Bloch sphere can be written $e^{-i \gamma \hat{n} \cdot \vec{\sigma} / 2}$, where $\vec{\sigma}$ is the vector of Pauli matrices $(X,Y,Z)$ (see Equation~4.8 of \cite{nielsen2002quantum}). Then, a single-qubit gate $U$ is equal to this rotation, up to a global phase $e^{i\alpha}$, i.e.,
\begin{align}
	U 
		&= e^{i\alpha} e^{-i \gamma \hat{n} \cdot \vec{\sigma} / 2} \nonumber \\
		&= e^{i\alpha} \left[ \cos \left( \frac{\gamma}{2} \right) I - i \sin \left( \frac{\gamma}{2} \right) \left( n_x X + n_y Y + n_z Z \right) \right] \label{eq:rotation_linear}.
\end{align}
Next, note \eqref{eq:rotation_linear} is a linear combination of $\{I, X, Y, Z\}$, i.e., it takes the form
\begin{equation}
	\label{eq:Pauli_basis}
	U = a I + b X + c Y + d Z,
\end{equation}
where
\begin{equation}
	\label{eq:rotation_coefficients}
	\begin{split}
		a &= e^{i\alpha} \cos \left( \frac{\gamma}{2} \right), \\
		b &= -i e^{i\alpha} \sin \left( \frac{\gamma}{2} \right) n_x, \\
		c &= -i e^{i\alpha} \sin \left( \frac{\gamma}{2} \right) n_y, \\
		d &= -i e^{i\alpha} \sin \left( \frac{\gamma}{2} \right) n_z. 
	\end{split}
\end{equation}
To find these coefficients, note for every $P,Q \in \{I, X, Y, Z\}$,
\[ \Tr(PQ) = \begin{cases}
	2, & P = Q, \\
	0, & P \ne Q, \\
\end{cases} \]
where $\Tr$ denotes the trace of a matrix, which is the sum of its diagonal elements. Then, multiplying \eqref{eq:Pauli_basis} by $I, X, Y, Z$ and taking the trace, we get
\begin{align*}
	\Tr(UI) &= a \Tr(I^2) + b \Tr(XI) + c \Tr(YI) + d \Tr(ZI) \\
		&= 2a, \\
	\Tr(UX) &= a \Tr(IX) + b \Tr(X^2) + c \Tr(YX) + d \Tr(ZX) \\
		&= 2b, \\
	\Tr(UY) &= a \Tr(IX) + b \Tr(XY) + c \Tr(Y^2) + d \Tr(ZY) \\
		&= 2c, \\
	\Tr(UZ) &= a \Tr(IX) + b \Tr(XZ) + c \Tr(YZ) + d \Tr(Z^2) \\
		&= 2d. 
\end{align*}
Thus,
\begin{equation}
	\label{eq:rotation_trace}
	\begin{split}
		a &= \frac{1}{2} \Tr(UI), \quad b = \frac{1}{2} \Tr(UX), \\
		c &= \frac{1}{2} \Tr(UY), \quad d = \frac{1}{2} \Tr(UZ).
	\end{split}
\end{equation}
Using these equations, we can find the coefficients $a$, $b$, $c$, and $d$ by multiplying the quantum gate $U$ by the appropriate matrix from $\{I, X, Y, Z\}$, taking the trace, and dividing by 2. Then, we can use \eqref{eq:rotation_coefficients} to find $\alpha$, $\gamma$, and $\hat{n} = (n_x, n_y, n_z)$.

For example, if
\[ U = HTHT, \]
then using \eqref{eq:rotation_trace},
\[ a = \frac{1}{2} \Tr(UI) = e^{i\pi/4} \frac{2+\sqrt{2}}{4}. \]
Comparing this to \eqref{eq:rotation_coefficients},
\begin{equation}
    \label{eq:rotation_angles}
    \alpha = \frac{\pi}{4}, \quad \gamma = 2 \cos^{-1} \left( \frac{2+\sqrt{2}}{4} \right).
\end{equation}
For the axis, combining \eqref{eq:rotation_coefficients} and \eqref{eq:rotation_trace}, we get
\begin{equation}
    \label{eq:rotation_axes}
    \begin{split}
        n_x &= \frac{\frac{1}{2}\Tr(UX)}{-i e^{i\alpha} \sin(\gamma/2)} = \frac{1}{\sqrt{5-2\sqrt{2}}}, \\
	    n_y &= \frac{\frac{1}{2}\Tr(UY)}{-i e^{i\alpha} \sin(\gamma/2)} = \frac{1-\sqrt{2}}{\sqrt{5-2\sqrt{2}}}, \\
	    n_z &= \frac{\frac{1}{2}\Tr(UZ)}{-i e^{i\alpha} \sin(\gamma/2)} = \frac{1}{\sqrt{5-2\sqrt{2}}}.
    \end{split}
\end{equation}
So we have found the angle and axis of rotation. Note in the introduction, the example was $(HT)^6$, so its angle of rotation is three times this, but the axis of rotation is the same.

Next, let us show how to express a single-qubit gate in the in the \zyz decomposition. We will do this by showing how to rewrite \eqref{eq:rotation_linear} in the \zyz decomposition. First, as a $2 \times 2$ matrix, \eqref{eq:rotation_linear} is
\begin{equation}
	\label{eq:rotation_matrix}
	U = e^{i\alpha} \begin{pmatrix}
		\cos \left( \frac{\gamma}{2} \right) - i n_z \sin \left( \frac{\gamma}{2} \right) & (-n_y - i n_x) \sin \left( \frac{\gamma}{2} \right) \\
		(n_y - i n_x) \sin \left( \frac{\gamma}{2} \right) & \cos \left( \frac{\gamma}{2} \right) + i n_z \sin \left( \frac{\gamma}{2} \right) \\
	\end{pmatrix}.
\end{equation}
Now, the \zyz decomposition (see Theorem 4.1 of \cite{nielsen2002quantum}) says we can write a single-qubit unitary as
\begin{align*}
	U
		&= e^{i\alpha'} R_z(\phi) R_y(\theta) R_z(\lambda) \\
		&= e^{i(\alpha'-\phi/2-\lambda/2)} \begin{pmatrix}
			\cos \left( \frac{\theta}{2} \right) & -e^{i\lambda} \sin \left( \frac{\theta}{2} \right) \\
			e^{i\phi} \sin \left( \frac{\theta}{2} \right) & e^{i(\phi+\lambda)} \cos \left( \frac{\theta}{2} \right) \\
	\end{pmatrix}.
\end{align*}
Renaming $\alpha' - \phi/2 - \lambda/2 \to \beta$, we get
\begin{equation}
	\label{eq:ZYZ_matrix}
	U = e^{i\beta} \begin{pmatrix}
		\cos \left( \frac{\theta}{2} \right) & -e^{i\lambda} \sin \left( \frac{\theta}{2} \right) \\
		e^{i\phi} \sin \left( \frac{\theta}{2} \right) & e^{i(\phi+\lambda)} \cos \left( \frac{\theta}{2} \right) \\
	\end{pmatrix}.
\end{equation}
Let us compare different elements of {\eqref{eq:rotation_matrix} and \eqref{eq:ZYZ_matrix}}. Starting with the top-left of each matrix, we get 
\[ e^{i\alpha} \left[ \cos \left( \frac{\gamma}{2} \right) - i n_z \sin \left( \frac{\gamma}{2} \right) \right] = e^{i\beta} \cos \left( \frac{\theta}{2} \right). \]
Dividing by $e^{i\alpha}$, this becomes
\[ \cos \left( \frac{\gamma}{2} \right) - i n_z \sin \left( \frac{\gamma}{2} \right) = e^{i(\beta-\alpha)} \cos \left( \frac{\theta}{2} \right). \]
Using Euler's formula, this becomes
\begin{align*}
	\cos \left( \frac{\gamma}{2} \right) - i n_z \sin \left( \frac{\gamma}{2} \right) 
		&= \cos(\beta-\alpha) \cos \left( \frac{\theta}{2} \right) \\
		&\quad + i\sin(\beta-\alpha) \cos \left( \frac{\theta}{2} \right).
\end{align*}
Matching the real and imaginary parts,
\begin{equation}
	\label{eq:ZYZ_re_im}
	\begin{split}
		&\cos \left( \frac{\gamma}{2} \right) = \cos(\beta-\alpha) \cos \left( \frac{\theta}{2} \right), \\
		&-n_z \sin \left( \frac{\gamma}{2} \right) = \sin(\beta-\alpha) \cos \left( \frac{\theta}{2} \right).
	\end{split}
\end{equation}
Dividing the second equation by the first,
\[ -n_z \tan \left( \frac{\gamma}{2} \right) = \tan(\beta-\alpha). \]
Solving for $\beta$,
\begin{gather}
	\tan^{-1} \left( -n_z \tan \frac{\gamma}{2} \right) = \beta - \alpha \nonumber \\
	-\tan^{-1} \left( n_z \tan \frac{\gamma}{2} \right) = \beta - \alpha \nonumber \\
	\beta = \alpha - \tan^{-1} \left( n_z \tan \frac{\gamma}{2} \right). \label{eq:beta}
\end{gather}

This lets us find the phase $\beta$. Next, back to \eqref{eq:ZYZ_re_im}, if we add the two equations, we get
\begin{align*}
	&\cos(\beta-\alpha) \cos \left( \frac{\theta}{2} \right) + \sin(\beta-\alpha) \cos \left( \frac{\theta}{2} \right) \\
	&\quad = \cos \left( \frac{\gamma}{2} \right) -n_z \sin \left( \frac{\gamma}{2} \right).
\end{align*}
Solving for $\theta$, we get
\begin{equation}
	\label{eq:theta}
	\theta = 2 \cos^{-1} \left( \frac{\cos \left( \frac{\gamma}{2} \right) -n_z \sin \left( \frac{\gamma}{2} \right)}{\sin(\beta-\alpha) + \cos(\beta-\alpha)} \right).
\end{equation}
Next, let us equate the top-right of \eqref{eq:rotation_matrix} and \eqref{eq:ZYZ_matrix}:
\[ e^{i\alpha} (-n_y - i n_x) \sin \left( \frac{\gamma}{2} \right) = -e^{i(\beta+\lambda)} \sin \left( \frac{\theta}{2} \right). \]
Dividing by $e^{i\alpha}$, we get
\[ (-n_y - i n_x) \sin \left( \frac{\gamma}{2} \right) = -e^{i(\beta+\lambda-\alpha)} \sin \left( \frac{\theta}{2} \right). \]
Using Euler's formula and matching the real and imaginary parts, we get
\begin{gather*}
	n_y \sin \left( \frac{\gamma}{2} \right) = \cos(\beta + \lambda - \alpha) \sin \left( \frac{\theta}{2} \right), \\
	n_x \sin \left( \frac{\gamma}{2} \right) = \sin(\beta + \lambda - \alpha) \sin \left( \frac{\theta}{2} \right).
\end{gather*}
Dividing these two equations, we get
\[ \frac{n_x}{n_y} = \tan(\beta + \lambda - \alpha). \]
Solving for $\lambda$,
\begin{equation}
	\label{eq:lambda}
	\lambda = \tan^{-1} \left( \frac{n_x}{n_y} \right) - \beta + \alpha.
\end{equation}
Now, let us equate the bottom-left of \eqref{eq:rotation_matrix} and \eqref{eq:ZYZ_matrix}:
\[ e^{i\alpha} (n_y - i n_x) \sin \left( \frac{\gamma}{2} \right) = e^{i(\beta+\phi)} \sin \left( \frac{\theta}{2} \right). \]
Dividing by $e^{i\alpha}$, we get
\[ (n_y - i n_x) \sin \left( \frac{\gamma}{2} \right) = e^{i(\beta+\phi-\alpha)} \sin \left( \frac{\theta}{2} \right) \]
Again using Euler's formula and matching the real and imaginary parts,
\begin{gather*}
	n_y \sin \left( \frac{\gamma}{2} \right) = \cos(\beta + \phi - \alpha) \sin \left( \frac{\theta}{2} \right), \\
	n_x \sin \left( \frac{\gamma}{2} \right) = -\sin(\beta + \phi - \alpha) \sin \left( \frac{\theta}{2} \right).
\end{gather*}
Dividing these,
\[ \frac{-n_x}{n_y} = \tan(\beta + \phi - \alpha). \]
Solving for $\phi$,
\begin{equation}
	\label{eq:phi}
	\phi= \tan^{-1} \left( \frac{-n_x}{n_y} \right) - \beta + \alpha.
\end{equation}
Thus, using \eqref{eq:theta}, \eqref{eq:lambda}, and \eqref{eq:phi}, we can find the angles to express any single-qubit gate in the \zyz decomposition.

For example, let us find the \zyz decomposition for
\[ (HT)^{6} = (HTHT)^{3}, \]
which we stated in Section II. Using our previous results for $HTHT$, $(HT)^6$ is a rotation about the same axis as \eqref{eq:rotation_axes}, but its global phase $e^{i\alpha}$ and angle of rotation $\gamma$ are three times what was given in \eqref{eq:rotation_angles}. That is, for $(HT)^{6}$  
\[ \alpha = \frac{3\pi}{4}, \quad \gamma = 6 \cos^{-1} \left( \frac{2+\sqrt{2}}{4} \right). \]
Plugging these into \eqref{eq:beta}, we get
\[ \beta = \frac{122\pi}{100}. \]
Then, we plug into \eqref{eq:theta}, \eqref{eq:phi}, and \eqref{eq:lambda} to get
\[ \theta = \frac{37\pi}{25}, \quad \phi = \frac{191\pi}{100}, \quad \lambda = \frac{29\pi}{25}. \]
These are the angles given in Section II.


\section{\label{appendix:addition}Proof of Addition Circuit}

In this appendix, we prove the behavior of the quantum addition circuit in \fref{q_addition_circuit}.
\begin{widetext}
\begin{align}
	|000000\rangle
	&\xrightarrow{X_{3}} |001000\rangle \label{first_eqn} \\
        &\xrightarrow{X_{5}} |001010\rangle \nonumber \\
	&\xrightarrow{H_{2}} \frac{1}{\sqrt{2}} \left( |001010\rangle + |011010\rangle \right) \nonumber \\
	&\xrightarrow{H_{6}} \frac{1}{2}\left( |001010\rangle + |001011\rangle + |011010 \rangle + |011011\rangle \right) \label{barrier_1}\\
        &\xrightarrow{H_{4}} \frac{1}{2\sqrt{2}}\left( |001010\rangle + |001011 \rangle + |001110\rangle + |001111\rangle + |011010\rangle + |011011 \rangle + |011110\rangle + |011111\rangle \right) \nonumber \\
        &\xrightarrow{C(S_{5,4})} \frac{1}{2\sqrt{2}}\left( |001010\rangle + |001011 \rangle + i|001110\rangle + i|001111\rangle + |011010\rangle + |011011 \rangle + i|011110\rangle + i|011111\rangle \right) \nonumber \\
        &\xrightarrow{C(T_{6,4})} \frac{1}{2\sqrt{2}} \big( |001010\rangle + |001011 \rangle + i|001110\rangle +  ie^{i\pi/4}|001111\rangle + |011010\rangle + |011011 \rangle + i|011110\rangle \nonumber \\
	&\quad\quad\quad\quad\quad\quad + ie^{i\pi/4}|011111\rangle \big) \nonumber \\
        &\xrightarrow{H_{5}} \frac{1}{4}\big( |001000\rangle + |001001\rangle  - |001010\rangle - |001011\rangle + i|001100 \rangle + ie^{i\pi/4}|001101 \rangle - i|001110\rangle  \nonumber \\
        &\quad\quad\quad - ie^{i\pi/4}|001111\rangle + |011000\rangle + |011001\rangle - |011010\rangle - |011011\rangle + i|011100 \rangle + ie^{i\pi/4}|011101 \rangle \nonumber \\
        &\quad\quad\quad - i|011110\rangle  - ie^{i\pi/4}|011111\rangle) \nonumber \\
        &\xrightarrow{C(S_{6,5})} \frac{1}{4}( |001000\rangle + |001001\rangle - |001010\rangle - i|001011\rangle + i|001100 \rangle + ie^{i\pi/4}|001101 \rangle - i|001110\rangle \nonumber \\
        &\quad\quad\quad\quad\quad+ e^{i\pi/4}|001111\rangle + |011000\rangle + |011001\rangle - |011010\rangle - i|011011\rangle + i|011100 \rangle + ie^{i\pi/4}|011101\rangle \nonumber\\
        &\quad\quad\quad\quad\quad- i|011110\rangle + e^{i\pi/4}|011111\rangle\big) \nonumber \\
        &\xrightarrow{H_{6}} \frac{1}{4\sqrt{2}}\big[ 2|001000\rangle -(1+i)|001010\rangle- (1-i)|001011 \rangle + i(1+e^{i\pi/4})|001100 \rangle+  i(1-e^{i\pi/4})|001101\rangle \label{barrier_2} \\
        &\quad\quad\quad\quad\quad+(e^{i\pi/4}-i)|001110 \rangle -(i+e^{i\pi/4})|001111\rangle+ 2|011000\rangle - (1+i) |011010 \rangle -(1-i)|011011\rangle\nonumber \\
        &\quad\quad\quad\quad\quad+ i(1+e^{i\pi/4})|011100 \rangle + i(1-e^{i\pi/4})|011101 \rangle  - (i-e^{i\pi/4})|011110 \rangle -(i+e^{i\pi/4})|011111\rangle \big] \nonumber \\
        &\xrightarrow{C(Z_{1,4})} \frac{1}{4\sqrt{2}}\big[ 2|001000\rangle -(1+i)|001010\rangle- (1-i)|001011 \rangle + i(1+e^{i\pi/4})|001100 \rangle+  i(1-e^{i\pi/4})|001101\rangle \nonumber \\
        &\quad\quad\quad\quad\quad\quad +(e^{i\pi/4}-i)|001110 \rangle -(i+e^{i\pi/4})|001111\rangle+ 2|011000\rangle - (1+i) |011010 \rangle -(1-i)|011011\rangle\nonumber \\
        &\quad\quad\quad\quad\quad\quad + i(1+e^{i\pi/4})|011100 \rangle + i(1-e^{i\pi/4})|011101 \rangle  - (i-e^{i\pi/4})|011110 \rangle -(i+e^{i\pi/4})|011111\rangle \big] \nonumber \\
        &\xrightarrow{C(S_{2,4})} \frac{1}{4\sqrt{2}}\big[2|001000\rangle -(1+i)|001010\rangle- (1-i)|001011 \rangle + i(1+e^{i\pi/4})|001100 \rangle+  i(1-e^{i\pi/4})|001101\rangle \nonumber \\
        &\quad\quad\quad\quad\quad\quad+(e^{i\pi/4}-i)|001110 \rangle   -(i+e^{i\pi/4})|001111\rangle + 2|011000\rangle - (1+i) |011010 \rangle -(1-i)|011011\rangle \nonumber \\
        &\quad\quad\quad\quad\quad\quad- (1+e^{i\pi/4})|011100 \rangle + (e^{i\pi/4}-1)|011101 \rangle  + (1+ie^{i\pi/4})|011110 \rangle +(1-ie^{i\pi/4})|011111\rangle \big]  \nonumber \\
        &\xrightarrow{C(T_{3,4})} \frac{1}{4\sqrt{2}}\big[2|001000\rangle -(1+i)|001010\rangle - (1-i)|001011 \rangle + (ie^{i\pi/4}-1)|001100 \rangle+  (1+ie^{i\pi/4})|001101\rangle  \nonumber \\
        &\quad\quad\quad\quad\quad\quad +i(1-e^{i\pi/4})|001110 \rangle  -i(1+e^{i\pi/4})|001111\rangle + 2|011000\rangle - (1+i) |011010 \rangle -(1-i)|011011\rangle \nonumber \\
        &\quad\quad\quad\quad\quad\quad - (i+e^{i\pi/4})|011100 \rangle + (i-e^{i\pi/4})|011101 \rangle  + (e^{i\pi/4}-1)|011110 \rangle +(1+e^{i\pi/4})|011111\rangle \big]  \nonumber \\
        &\xrightarrow{C(Z_{2,5})} \frac{1}{4\sqrt{2}}\big[2|001000\rangle -(1+i)|001010\rangle- (1-i)|001011 \rangle + (ie^{i\pi/4}-1)|001100 \rangle+  (1+ie^{i\pi/4})|001101\rangle  \nonumber \\
        &\quad\quad\quad\quad\quad\quad +i(1-e^{i\pi/4})|001110 \rangle -i(1+e^{i\pi/4})|001111\rangle + 2|011000\rangle + (1+i) |011010 \rangle + (1-i)|011011\rangle \nonumber \\
        &\quad\quad\quad\quad\quad\quad - (i+e^{i\pi/4})|011100 \rangle + (i-e^{i\pi/4})|011101 \rangle  - (e^{i\pi/4}-1)|011110 \rangle - (1+e^{i\pi/4})|011111\rangle \big]  \nonumber \\
        &\xrightarrow{C(S_{3,5})} \frac{1}{4\sqrt{2}}\big[2|001000\rangle + (1-i)|001010\rangle - (1+i)|001011 \rangle + (ie^{i\pi/4}-1)|001100 \rangle+  (1+ie^{i\pi/4})|001101\rangle \nonumber \\
        &\quad\quad\quad\quad\quad\quad - (1-e^{i\pi/4})|001110 \rangle  + (1+e^{i\pi/4})|001111\rangle + 2|011000\rangle + (i-1) |011010 \rangle + (1+i)|011011\rangle \nonumber \\
        &\quad\quad\quad\quad\quad\quad - (i+e^{i\pi/4})|011100 \rangle + (i-e^{i\pi/4})|011101 \rangle  - i(e^{i\pi/4}-1)|011110 \rangle - i(1+e^{i\pi/4})|011111\rangle \big]  \nonumber \\
	&\xrightarrow{C(Z_{3,6})}  \frac{1}{4\sqrt{2}}\big[2|001000\rangle + (1-i)|001010\rangle + (1+i)|001011 \rangle + (ie^{i\pi/4}-1)|001100 \rangle \label{barrier_3} \\
        &\quad\quad\quad\quad\quad\quad - (1+ie^{i\pi/4})|001101\rangle - (1-e^{i\pi/4})|001110 \rangle  - (1+e^{i\pi/4})|001111\rangle + 2|011000\rangle \nonumber \\
        &\quad\quad\quad\quad\quad\quad + (i-1) |011010 \rangle - (1+i)|011011\rangle - (i+e^{i\pi/4})|011100 \rangle - (i-e^{i\pi/4})|011101 \rangle \nonumber \\
	&\quad\quad\quad\quad\quad\quad - i(e^{i\pi/4}-1)|011110 \rangle + i(1+e^{i\pi/4})|011111\rangle \big]  \nonumber \\
        &\xrightarrow{H_{6}} \frac{1}{4}\big( |001000\rangle + |001001\rangle +|001010 \rangle -i|001011\rangle -|001100\rangle + ie^{i\pi/4}|001101\rangle- |001110 \rangle + e^{i\pi/4}|001111\rangle \nonumber \\
        &\quad\quad\quad + |011000 \rangle + |011001\rangle  -|011010\rangle + i|011011\rangle - i|011100 \rangle - e^{i\pi/4}|011101\rangle + i|011110\rangle \nonumber \\
        &\quad\quad\quad- ie^{i\pi/4}|011111\rangle    \big) \nonumber \\
        &\xrightarrow{C(S_{6,5}^{\dagger})} \frac{1}{4}\big( |001000\rangle + |001001\rangle +|001010 \rangle -|001011\rangle -|001100\rangle + ie^{i\pi/4}|001101\rangle - |001110 \rangle \nonumber \\
        &\quad\quad\quad\quad\quad - ie^{i\pi/4}|001111\rangle  + |011000 \rangle + |011001\rangle  -|011010\rangle + |011011\rangle - i|011100 \rangle - e^{i\pi/4}|011101\rangle \nonumber \\
        &\quad\quad\quad\quad\quad + i|011110\rangle  - e^{i\pi/4}|011111\rangle \big) \nonumber \\
        &\xrightarrow{H_{5}} \frac{1}{2\sqrt{2}}\big( |001000\rangle + |001011\rangle - |001100 \rangle + ie^{i\pi/4}|001111\rangle + |011001\rangle + |011010 \rangle - e^{i\pi/4}|011101\rangle  \nonumber \\
        &\quad\quad\quad\quad-i|011110\rangle   \big) \nonumber \\
        &\xrightarrow{C(T_{6,4} ^{\dagger})} \frac{1}{2\sqrt{2}}\big( |001000\rangle + |001011\rangle - |001100 \rangle + i|001111\rangle + |011001\rangle + |011010 \rangle - |011101\rangle -i|011110\rangle   \big) \nonumber \\
        &\xrightarrow{C(S_{5,4} ^{\dagger})} \frac{1}{2\sqrt{2}}\big( |001000\rangle + |001011\rangle - |001100 \rangle + |001111\rangle + |011001\rangle + |011010 \rangle - |011101\rangle -|011110\rangle \big) \nonumber \\
        &\xrightarrow{H_{4}} \frac{1}{2}\big( |001011\rangle + |001100\rangle +  |011101 \rangle + |011110\rangle \big) \label{final_eqn}.
\end{align}
\end{widetext}
In the above equation, the left-hand side of \eqref{first_eqn} is the initial state of the quantum circuit in \fref{q_addition_circuit}, which has four sections. \eqref{barrier_1} is the state after the first section, \eqref{barrier_2} is the state after the second section, \eqref{barrier_3} is the state after the third section, and \eqref{final_eqn} is the state after the fourth section, and it is the final state of the circuit.


\bibliography{bibliography}
\bibliographystyle{aps}

\end{document}